# On an attempt to resolve the EPR-Bell paradox via Reichenbachian concept of common cause

**László E. Szabó***

*Theoretical Physics Research Group of HAS*
*Department of History and Philosophy of Science*
*Eötvös University, Budapest*

Abstract

Reichenbach's Common Cause Principle claims that if there is correlation between two events and none of them is directly causally influenced by the other, then there must exist a third event that can, as a common cause, account for the correlation. The EPR-Bell paradox consists in the problem that we observe correlations between spatially separated events in the EPR-experiments, which do not admit common-cause-type explanation; and it must therefore be inevitably concluded, that, contrary to relativity theory, in the realm of quantum physics there exists action at a distance, or at least superluminal causal propagation is possible; that is, either relativity theory or Reichenbach's common cause principle fails.

By means of closer analyses of the concept of common cause and a more precise reformulation of the EPR experimental scenario, I will sharpen the conclusion we can draw from the violation of Bell's inequalities. It will be explicitly shown that the correlations we encounter in the EPR experiment could have common causes; that is, Reichenbach's Common Cause Principle does not fail in quantum mechanics. Moreover, these common causes satisfy the locality conditions usually required.

In the Revised Version of the paper I added a Postscript from which it turns out that the solution such obtained is, contrary to the original title, incomplete. It turns out that a new problem arises: some *combinations* of the common cause events do statistically correlate with the measurement operations.

## *Introduction*

**1.** In his last work on 'Direction of Time'[1] Hans Reichenbach was the first in the philosophical literature, who analyzed what causality means in a stochastic theory. He formulated the causal relations of events by means of the statistical correlations among them. Two events are correlated if the following holds for their probabilities:

---

* E-mail: szabol@caesar.elte.hu

[1] Reichenbach, H., *The Direction of Time*, University of California Press, Los Angeles, 1956.

$$\Delta_p(AB) = p(AB) - p(A)p(B) \neq 0$$

Seeing correlation between two events $A$ and $B$, one can imagine two kinds of explanation: 1) the occurrence of one event is directly influenced by the occurrence of the other (so called *direct correlation*), or 2) the correlation is explained by the existence of a third event *Z*, a *common cause*, which is directly correlated to both *A* and *B*. In this case we say that the correlation is a *common cause correlation.*

Following Reichenbach[2], we give the following definition for a common cause:

*Let $A$ and $B$ be two correlated events, $\Delta(AB) \neq 0$. An event $Z$ is called common cause, if*

$$Z \neq A, B \tag{1}$$

$$\begin{aligned} &\text{if } Z \subset A \text{ or } Z \supset A \text{ then } p(Z) \neq p(A) \\ &\text{if } Z \subset B \text{ or } Z \supset B \text{ then } p(Z) \neq p(B) \end{aligned} \tag{2}$$

$$p(AB \mid Z) = p(A \mid Z)\, p(B \mid Z) \tag{3}$$

$$p(AB \mid \overline{Z}) = p(A \mid \overline{Z})\, p(B \mid \overline{Z}) \tag{4}$$

$$\begin{aligned} &\begin{Bmatrix} p(AZ) > p(A)p(Z) \\ p(BZ) > p(B)p(Z) \end{Bmatrix} \text{ or } \begin{Bmatrix} p(AZ) < p(A)p(Z) \\ p(BZ) < p(B)p(Z) \end{Bmatrix} \quad \text{if } \Delta(AB) > 0 \\ &\begin{Bmatrix} p(AZ) > p(A)p(Z) \\ p(BZ) < p(B)p(Z) \end{Bmatrix} \text{ or } \begin{Bmatrix} p(AZ) < p(A)p(Z) \\ p(BZ) > p(B)p(Z) \end{Bmatrix} \quad \text{if } \Delta(AB) < 0 \end{aligned} \tag{5}$$

I need to mention that (1) and (2) are not originally contained in Reichenbach's definition. That is to say, the original definition includes the direct causal relation too, as well as the possibility that the common cause differs from *A* and *B* only in a probability-0 event, which possibility we want to exclude, of course. (5) also contains some slight changes, namely that, unlike Reichenbach, we do not want to restrict ourselves to positive correlations. Only the so-called "screening off" properties (3) and (4) from the definition of common cause need some explanation: If we restrict the statistical ensemble to the sub-ensemble, in which the occurrence (or non-occurrence respectively) of the common cause event is fixed, then the correlation disappears.

**2.** Reichenbach's Common Cause Principle (CCP) is this: If two evens are correlated and one can exclude the possibility of direct causal relationship between them, then there must exist an event satisfying all the conditions required in the above definition of common cause. This is a very strong metaphysical claim about the causal structure of our world, however it is completely compatible with our intuition.

We cannot assume in general, that for each correlated pair of events there exist a suitable event in the original event algebra used for the modeling of the phenomena in question, such that it satisfies the definition of common cause. Of course, the existence of such an element in the original algebra is not required by the CCP, but it requires the existence of a common cause event *in reality*. However, if for each correlation there is a common cause in reality then we may with good reason to assume that the original event algebra is extendable in such a way that all of these common causes are contained in the extension. Otherwise, we would find ourselves in an extremely counterintuitive situation as observing events in the world, but about which we would not be able to speak with the logical connectives of the everyday language.[3] Thus, there would be an obvious strategy to prove the invalidity of the CCP, if someone shown up correlated events described by a probabilistic model which were not extendable with common causes for all correlations. This is typically the strategy of the EPR-Bell theorem aimed to prove the failure of the CCP in quantum mechanics.

[2]Ibid. p. 159.

[3]Cf. Dummett, M., *The Logical Basis of Metaphysics*, Duckworth, London, 1995, p.54.

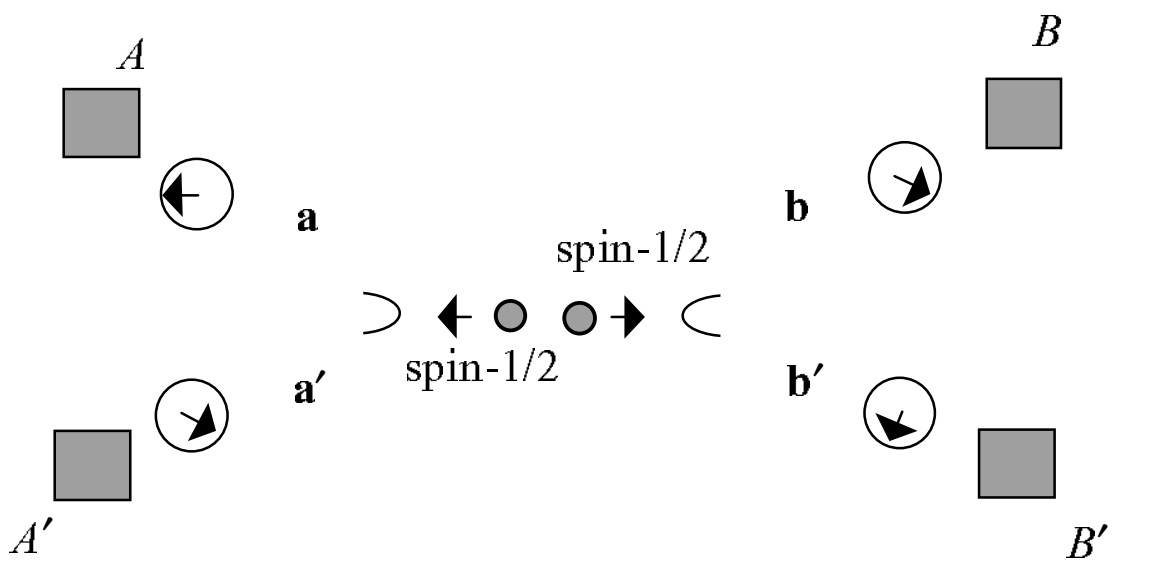

***Figure 1** The Aspect experiment with spin-$\frac{1}{2}$ particles*

**3.** **Figure 1** shows the well-known Aspect experiment. The four detectors detect the spin-up events in the spin-component measurements in directions $\mathbf{a}, \mathbf{a}'$ and $\mathbf{b}, \mathbf{b}'$. There are random switches (independent agents, if you want) choosing between the different possible measurements on both sides. Let $p(a), p(a')$ and $p(b), p(b')$ be *arbitrary* probabilities with which the different measurements are chosen. We can experience the following events in the experiment:

- $A$: "the spin of the left particle is *up* in direction $\mathbf{a}$" detector fires
- $A'$: "the spin of the left particle is *up* in direction $\mathbf{a}'$" detector fires
- $B$: "the spin of the left particle is *up* in direction $\mathbf{b}$" detector fires
- $B'$: "the spin of the left particle is *up* in direction $\mathbf{a}'$" detector fires
- $a$: the left switch chooses the $\mathbf{a}$-measurement
- $a'$: the left switch chooses the $\mathbf{a}'$-measurement
- $b$: the right switch chooses the $\mathbf{b}$-measurement
- $b'$: the right switch chooses the $\mathbf{b}'$-measurement

Let $\mathbf{a}, \mathbf{a}', \mathbf{b}, \mathbf{b}'$ be coplanar vectors such that $\sphericalangle(\mathbf{a}, \mathbf{a}') = \sphericalangle(\mathbf{a}', \mathbf{b}') = \sphericalangle(\mathbf{a}, \mathbf{b}') = 120^\circ$, and $\sphericalangle(\mathbf{a}', \mathbf{b}) = 0$. We observe the following relative frequencies in the experiment:

$$\begin{aligned} p(A) &= \frac{1}{2}p(a) & p(B) &= \frac{1}{2}p(b) \\ p(A') &= \frac{1}{2}p(a') & p(B') &= \frac{1}{2}p(b') \\ p(AB) &= \frac{3}{8}p(a)p(b) & p(A'B) &= 0 \\ p(AB') &= \frac{3}{8}p(a)p(b') & p(A'B') &= \frac{3}{8}p(a')p(b') \end{aligned} \tag{6}$$

Visibly, there are correlations among the outcomes of the measurements performed on the left and the right particles:

$$\begin{aligned} \Delta(AB) &= p(AB) - p(A)p(B) \neq 0 \\ \Delta(AB') &= p(AB') - p(A)p(B') \neq 0 \\ \Delta(A'B) &= p(A'B) - p(A')p(B) \neq 0 \\ \Delta(A'B') &= p(A'B') - p(A')p(B') \neq 0 \end{aligned} \tag{7}$$

In the ensuing sections of this paper I shall argue that, contrary to the standard view, each correlation observed here between spatially separated events does have a common cause in an appropriately extended event algebra. However, it is worth recalling first the train of thought how the Bell-inequalities are derived and how the failure of the CCP is usually argued.

**4.** The usual interpretation of the experimental data in (6) is the following: The conditional probability $p(A \mid a) \overset{def}{=} \frac{p(Aa)}{p(a)}$, for example, is regarded as the quantum mechanical probability, $\operatorname{tr}\left(\hat{W}\hat{A}\right)$, of a particular property represented by projector $\hat{A}$.[4] In this sense, the data in (6) are in accordance with the quantum mechanical prediction. The various versions of the EPR-Bell theorem apply just to these derived conditional probabilities, which are, according to the standard view, understood as the "quantum probabilities" of certain "properties" of the particles. Let us denote these probabilities as follows:

$$
\begin{array}{rlrl}
q(A) = p(A \mid a) = & \frac{1}{2} & q(AB) = p(AB \mid ab) = & \frac{3}{8} \\
q(A') = p(A' \mid a') = & \frac{1}{2} & q(A'B) = p(A'B \mid a'b) = & 0 \\
q(B) = p(B \mid b) = & \frac{1}{2} & q(AB') = p(AB' \mid ab') = & \frac{3}{8} \\
q(B') = p(B' \mid b') = & \frac{1}{2} & q(A'B') = p(A'B' \mid a'b') = & \frac{3}{8}
\end{array}
\tag{8}
$$

where $q(A)$ is interpreted as the "quantum probability" of that the left particle is of property "spin-up". The correlations in (7) can be expressed by using these quantum probabilities, too:

$$
\begin{array}{r}
\Delta_q(AB) = \Delta_q(AB') = \Delta_q(A'B') = \frac{1}{8} \\
\Delta_q(A'B) = -\frac{1}{4}
\end{array}
\tag{9}
$$

The question, as it is formulated almost everywhere in the literature of the EPR-Bell paradox, is this:

(*) DOES A COMMON CAUSE FOR THE CORRELATIONS IN (9) EXIST?

That is to say, is it possible to imagine an extension of the event algebra containing an event $Z$ satisfying conditions (1)-(5) with respect of the "quantum probabilities" defined in (8)?[5]

As I have pointed out in earlier papers[6], this question, in the above formulation, is meaningless. I am convinced that we must not ignore the fact that – just because of the violation of the $n = 4$ Pitowsky inequalities[7] – the numbers $q(A)$, $q(A')$, $q(B)$, $\ldots$ are not interpretable as relative frequencies and do not form a Kolmogorovian probability system that the Reichenbach axioms could apply to. There are no events – and in principle, there cannot exist events in reality the probabilities (relative frequencies) of which would be equal to "quantum probabilities" $q(A),\ q(A'),\ q(B),\ \ldots$[8] [9]. In other words, (*) is a question about the existence of common cause for correlations among *non-existing* events.

---

[4]Without going into the details of the quantum mechanical calculation, $\hat{W}$ and $\hat{A}$ are the following projectors of the "spin state space" $H^2 \otimes H^2$ : $\hat{W} = \hat{P}_{\operatorname{span}\left\{\sigma_{\mathbf{a}}^+ \otimes \sigma_{\mathbf{a}}^- - \sigma_{\mathbf{a}}^- \otimes \sigma_{\mathbf{a}}^+\right\}}$ and $\hat{A} = \hat{P}_{\operatorname{span}\left\{\sigma_{\mathbf{a}}^+ \otimes \sigma_{\mathbf{a}}^+, \sigma_{\mathbf{a}}^+ \otimes \sigma_{\mathbf{a}}^-\right\}}$.

[5]It is to be noted that the notion of common cause used in the stochastic hidden variable model is weaker than that of Reichenbach; the common cause defined in (1)-(5) is equivalent with the notion of two-valued hidden parameter: $\lambda = \lambda_1 \Leftrightarrow Z$, $\lambda = \lambda_2 \Leftrightarrow \bar{Z}$.

[6]Szabó, L. E., Is quantum mechanics compatible with a deterministic universe? Two interpretations of quantum probabilities, *Foundations of Physics Letters*, **8**, 421-440. (1995); Quantum Mechanics in an Entirely Deterministic Universe, *Int. J. Theor. Phys.*, **34**, 1751-1766. (1995)

[7]Pitowsky, I., *Quantum Probability — Quantum Logic* (Lecture Notes in Physics **321**), Springer, Berlin, (1989).

It will be important in the ensuing parts of this paper, that the $\mathrm{tr}\left(\hat{W}\hat{A}\right)$-like quantities should be correctly interpreted as conditional probabilities, and we will see how this divergence from the standard view implies important differences with respect to the validity of the CCP. First, I would like to show, however, that question (*) can be meaningfully reformulated in terms of conditional probabilities. Nevertheless, the final result will be the same: one can derive the Clause-Horne inequalities within the correct conceptual framework, too, and the violation of them, in a certain sense, excludes the existence of a common cause. It will be clarified soon in what sense.

## *No Common Common Cause*

**5.** Consider one after the other all the important facts about the EPR-Aspect experiment. 1) For example, the left particle emitted from the source is subjected to one of the measurement procedures $a$ or $a'$. Obviously, the outcome of the measurement is effected by the preceding measurement procedure, and this effect shows regularities described by quantum mechanics. 2) It is also obvious that, because of the spatial separation, the measurement outcome on the one side must be independent of the measurement operation performed on the other side. 3) The choices of the measurements are free, therefore the left and right measurement selections are statistically independent. 4) We can speak about correlations between events on the left and right hand sides only in case of the outcomes of the measurements. Assume there is an event that is a common cause for such a correlation. Of course, the probabilities of the measurement choices $p(a), p(a'), p(b), p(b')$ are entirely arbitrary. Consequently, 5) one can require that the common cause event be a common cause independently of the concrete values of probabilities $p(a), p(a'), p(b), p(b')$, and that the probability of the common cause be also independent from the probabilities of the measurement choices. 6) Also, because the choices of the measurements are free in the sense that there is no mysterious conspiracy between the things that determine the choices of the measurements and those that determine the outcomes, one can assume that the measurement choices are independent of the common cause. 7) For sake of simplicity, we can finally assume that one of the measurements is surely performed on the both hand sides.

These findings are partly read off from the empirical data (6) or they are straightforward consequences of the prohibition of superluminal causation.

The above requirements can be expressed in the following formulas:

$$
\begin{aligned}
p(X) &= p(X \mid x)\, p(x) = \mathrm{tr}\left(\hat{W}\hat{X}\right) p(x) \\
p(Y) &= p(Y \mid y)\, p(y) = \mathrm{tr}\left(\hat{W}\hat{Y}\right) p(y)
\end{aligned}
\tag{10}
$$

$$
p(xy) = p(x)\, p(y) \tag{11}
$$

$$
\begin{aligned}
p(Xy) &= p(X)\, p(y) \\
p(xY) &= p(x)\, p(Y)
\end{aligned}
\tag{12}
$$

[8] Szabó, L. E., Quantum structures do not exist in reality, *Int. J. of Theor. Phys.*, **37**, 449-456. (1998).

[9] It worth to mention that already von Neumann was aware of that it is impossible to interpret quantum probabilities as relative frequencies. See Rédei, M., Why John von Neumann did not like the Hilbert space formalism of quantum mechanics (and what he liked instead), *Studies in the History and Philosophy of Modern Physics*, **27**, 493-510. (1996).

$$p\left(xZ_{XY}\right)=p(x)\,p\left(Z_{XY}\right)$$
$$p\left(yZ_{XY}\right)=p(y)\,p\left(Z_{XY}\right) \tag{13}$$

$$p(a)+p\left(a'\right)=p(b)+p\left(b'\right)=1 \tag{14}$$

$$Z_{XY}\neq X,Y \tag{15}$$

$$\text{if } Z\subset X \text{ or } Z\supset X \text{ then } p\left(Z_{XY}\right)\neq p(X)$$
$$\text{if } Z\subset Y \text{ or } Z\supset Y \text{ then } p\left(Z_{XY}\right)\neq p(Y) \tag{16}$$

$$p\left(XY\mid Z_{XY}\right)=p\left(X\mid Z_{XY}\right)p\left(Y\mid Z_{XY}\right) \tag{17}$$

$$p\left(XY\mid \bar{Z}_{XY}\right)=p\left(X\mid \bar{Z}_{XY}\right)p\left(Y\mid \bar{Z}_{XY}\right) \tag{18}$$

$$\begin{array}{l}\left\{\begin{array}{l}p\left(XZ_{XY}\right)>p(X)\,p\left(Z_{XY}\right)\\ p\left(YZ_{XY}\right)>p(Y)\,p\left(Z_{XY}\right)\end{array}\right\} \text{ or } \left\{\begin{array}{l}p\left(XZ_{XY}\right)<p(X)\,p\left(Z_{XY}\right)\\ p\left(YZ_{XY}\right)<p(Y)\,p\left(Z_{XY}\right)\end{array}\right\} \text{ if } \Delta(XY)>0\\ \left\{\begin{array}{l}p\left(XZ_{XY}\right)>p(X)\,p\left(Z_{XY}\right)\\ p\left(YZ_{XY}\right)<p(Y)\,p\left(Z_{XY}\right)\end{array}\right\} \text{ or } \left\{\begin{array}{l}p\left(XZ_{XY}\right)<p(X)\,p\left(Z_{XY}\right)\\ p\left(YZ_{XY}\right)>p(Y)\,p\left(Z_{XY}\right)\end{array}\right\} \text{ if } \Delta(XY)<0\end{array} \tag{19}$$

where $Z_{XY}$ denotes the common cause for correlation $\Delta(XY)\neq 0$, and $X=A,A'$; $Y=B,B'$; $x=a,a'$; $y=b,b'$.

**6.** Since our aim is to give a correct derivation of a Bell-type inequality, we must explicitly stipulate one more assumption which has, as a tacit assumption, always been present in the literature of Bell inequalities, and which I willfully left vague in point **4.** We are looking for common causes of four different correlations in the EPR-Aspect experiment. It is tacitly assumed in the various derivations of Bell inequalities, that the four correlated pairs of events have one single *common* common cause[10]. This assumption is, however, completely unjustified, because one can easily show classical physical examples with correlations, for which there is no common common cause, but which admit separate common causes. Nevertheless, because our aim is to derive the Clauser-Horne inequality, we must explicitly stipulate that the common causes for correlated pairs of events $(A,B)$, $\left(A,B'\right)$, $\left(A',B\right)$ and $\left(A',B'\right)$ coincide:

$$Z_{AB}=Z_{AB'}=Z_{A'B}=Z_{A'B'}=Z \tag{20}$$

that is, $Z$ is a common common cause.

Now, we sharpen the question about common cause by modifying question (*) in the following way:

(**) DOES A COMMON COMMON CAUSE FOR THE CORRELATIONS IN (7) EXIST, SATISFYING CONDITIONS (10)-(20)?

In this form, the question refers only to real events and real probabilities understandable as relative frequencies.

**7.** Suppose $Z$ is such an event. Then, it follows from (10)-(20) that

[10]See Belnap, N. and Szabó, L. E., Branching Space-time analysis of the GHZ theorem, *Foundations of Physics*, **26**, 989-1002. (1996), where we first pointed out the importance of the tacit assumption in the Bell theorem that the common cause is understood as a common common cause.

$$
\begin{aligned}
p(XY \mid xy) &= \frac{p(XYxy)}{p(xy)} = \frac{p(XYxyZ)}{p(xy)} + \frac{p(XYxy\bar{Z})}{p(xy)} \\
&= \frac{p(XYZ)}{p(xy)} + \frac{p(XY\bar{Z})}{p(xy)} = \frac{p(XY \mid Z)}{p(xy)} p(Z) + \frac{p(XY \mid \bar{Z})}{p(xy)} p(\bar{Z}) \\
&= \frac{p(X \mid Z)}{p(x)} \frac{p(Y \mid Z)}{p(y)} p(Z) + \frac{p(X \mid \bar{Z})}{p(x)} \frac{p(Y \mid \bar{Z})}{p(y)} p(\bar{Z}) \\
&= \frac{p(XxZ)}{p(x)p(Z)} \frac{p(YyZ)}{p(y)p(Z)} p(Z) + \frac{p(Xx\bar{Z})}{p(x)p(\bar{Z})} \frac{p(Yy\bar{Z})}{p(y)p(\bar{Z})} p(\bar{Z}) \\
&= p(X \mid xZ)\, p(Y \mid yZ)\, p(Z) + p(X \mid x\bar{Z})\, p(Y \mid y\bar{Z})\, p(\bar{Z})
\end{aligned}
$$

Similarly,

$$
\begin{aligned}
p(X \mid x) &= p(X \mid xZ)\, p(Z) + p(X \mid x\bar{Z})\, p(\bar{Z}) \\
p(Y \mid y) &= p(Y \mid yZ)\, p(Z) + p(Y \mid y\bar{Z})\, p(\bar{Z})
\end{aligned}
$$

where $X = A, A'$; $Y = B, B'$; $x = a, a'$ and $y = b, b'$. Inequality

$$-1 \leq uv + uv' - u'v + u'v' - u - v' \leq 0$$

holds for arbitrary real numbers $u, u', v, v' \in [0,1]$. Therefore,

$$
\begin{aligned}
-1 \leq\ & p(AB \mid abZ) + p(AB' \mid ab'Z) - p(A'B \mid a'bZ) \\
& + p(A'B' \mid a'b'Z) - p(A \mid aZ) - p(B' \mid b'Z) \leq 0
\end{aligned}
$$

$$
\begin{aligned}
-1 \leq\ & p(AB \mid ab\bar{Z}) + p(AB' \mid ab'\bar{Z}) - p(A'B \mid a'b\bar{Z}) \\
& + p(A'B' \mid a'b'\bar{Z}) - p(A \mid a\bar{Z}) - p(B' \mid b'\bar{Z}) \leq 0
\end{aligned}
$$

Multiplying by $p(Z)$ and $p(\bar{Z})$ respectively, and summing up the two inequalities, we get the well-known Clauser-Horne inequality:

$$
\begin{aligned}
-1 \leq\ & p(AB \mid ab) + p(AB' \mid ab') - p(A'B \mid a'b) \\
& + p(A'B' \mid a'b') - p(A \mid a) - p(B' \mid b') \leq 0
\end{aligned}
\tag{21}
$$

Substituting the corresponding data from (6), we find that (21) fails:

$$\frac{3}{8} + \frac{3}{8} - 0 + \frac{3}{8} - \frac{1}{2} - \frac{1}{2} = \frac{1}{8} > 0$$

Consequently, there cannot exist a common common cause for the correlations in (7), that would satisfy conditions (10)-(20).

It is worth to make two remarks with respect of the above-derived Clauser-Horne inequality:

1) Inequality (21) is similar to, but not identical with the "Clauser-Horne" inequality derived by Pitowsky[11]. Namely, the above-derived inequality is the same as the original one of

[11]Ibid. p. 27.

Clauser and Horne[12]. The obvious difference is that the original inequality applies to conditional probabilities, while that of Pitowsky contains only absolute probabilities.

2) Pitowsky's "Clauser-Horne" inequality, just like the other Bell-type inequalities derived by him, is destined to indicate whether or not the "probabilities" in question are representable in a Kolmogorov probability model. At the same time, the original Clauser-Horne inequality indicates whether there can exist a common common cause for the considered correlations. All the probabilities in (6) can be accommodated in a Kolmogorovian theory,[13] without any problem, although there is no common common cause for the correlations. That is to say, the validity of the Kolmogorov axioms and that of the CCP are different problems.

## *Separate common causes exist*

**8.** As we have seen, the Clauser-Horne inequality can be derived within a conceptually correct framework, too, without referring to "quantum"/non-existing events and "quantum probabilities". We have also seen, that the relative frequencies measured in the EPR-Aspect experiment violate this inequality. The consequence we have drawn is that there cannot exist a common common cause for the correlations observed in the experiment.

As I have already mentioned in point **6.**, the assumption that the considered correlations have a *common* common cause is completely unjustified, since one can easily show classical physical examples with correlations, for which there is no common common cause, but which admit separate common causes. This of course raises the question whether there exist separate common causes for the correlations observed in the EPR-Aspect experiment. In the remaining part of the paper I shall prove that this is possible.

**9.** First, I define a finite event algebra with a probability measure, that is able to model the EPR-Aspect experiment. Then, I am going to show that the event algebra is extendable such that all correlations have a common cause.

Let the algebra in which the events of the EPR-Aspect experiment will be represented be the Boolean algebra $\mathcal{A} = 2^{\Omega}$, generated by the following set of atoms: $\Omega = \left\{u_1, u_2, \ldots u_{16}\right\}$ (see **Figure 2**). Denote $p_1, p_2, \ldots\ p_{16}$ the probabilities of the atoms. From conditions (10)-(19) there follow some obvious correlations that we do not want to deal with, since they do not mean any problem with respect to locality:

$$\Delta(A,a) \neq 0\ ;\ \Delta(A,a') \neq 0\ ;\ \Delta(A,'a) \neq 0; \Delta(A',a') \neq 0$$

$$\Delta(B,b) \neq 0; \Delta(B,b') \neq 0; \Delta(B',b) \neq 0; \Delta(B',b') \neq 0$$

$$\Delta(a,a') \neq 0; \Delta(b,b') \neq 0$$

[12] Clauser, J. F. and Shimony, A., "Bell's theorem: experimental tests and implications", *Rep. Prog. Phys.*, **41**, 1881. (1978).

[13] See Szabó, L. E., Is quantum mechanics compatible with a deterministic universe? Two interpretations of quantum probabilities, *Foundations of Physics Letters*, **8**, 421-440. (1995); Quantum Mechanics in an Entirely Deterministic Universe, *Int. J. Theor. Phys.*, **34**, 1751-1766. (1995); Quantum structures do not exist in reality, *Int. J. of Theor. Phys.*, **37**, 449-456. (1998); Bana, G. and Durt, T., Proof of Kolmogorovian Censorship, *Foundations of Physics*, **27**, 1355-1373. (1997); Gill, R. D., Critique of 'Elements of Quantum Probability', in Quantum Probability Communications XI, R. L. Hudson and J. M. Lindsay eds., World Scientific, Singapore, 1998.

These correlations are represented by the "logical structure" of the events. Following the notations in (8), from (6) and (10)-(19) we obtain the following equations:

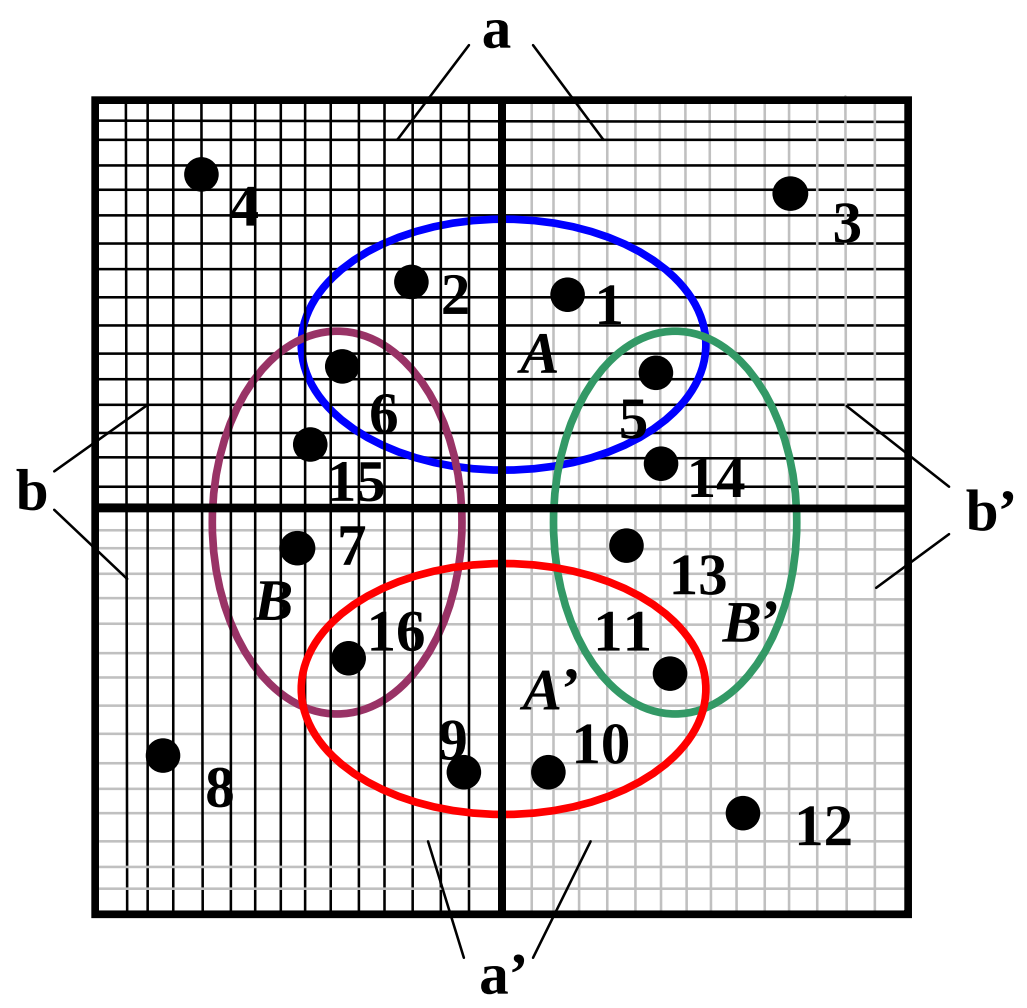

***Figure 2*** *The Boolean event algebra in which the events of the EPR-Aspect experiment are represented.*

$$p(b')q(B') = p_5 + p_{11} + p_{13} + p_{14}$$
$$p(a)p(b) = p_4 + p_2 + p_6 + p_{15}$$
$$p(a')p(b) = p_7 + p_8 + p_9 + p_{16}$$
$$p(a)p(b') = p_3 + p_1 + p_5 + p_{14}$$
$$p(a')p(b') = p_{12} + p_{10} + p_{11} + p_{13}$$
$$p(a)p(b)q(A) = p_2 + p_6$$
$$p(a')p(b)q(A') = p_9 + p_{16}$$
$$p(a)p(b')q(A) = p_1 + p_5$$
$$p(a')p(b')q(A') = p_{10} + p_{11}$$
$$p(a)p(b)q(B) = p_{15} + p_6$$
$$p(a')p(b)q(B) = p_7 + p_{16}$$
$$p(a)p(b')q(B') = p_5 + p_{14}$$
$$p(a)p(b')q(AB') = p_5$$
$$p(a)p(b)q(AB) = p_6$$
$$p(a')p(b')q(A'B') = p_{11}$$
$$p(a')p(b)q(B) = p_{16}$$

This is an easily soluble system of equations with 16 unknowns. As a solution, we obtain the following probabilities:

$$
\begin{aligned}
p_1 &= p(a)\,p(b')\left(q(A)-q(AB')\right)\\
p_2 &= p(a)\,p(b)\left(q(A)-q(AB)\right)\\
p_3 &= p(a)\,p(b')\left(1-q(A)-q(B')+q(AB')\right)\\
p_4 &= p(a)\,p(b)\left(1-q(A)-q(B)+q(AB)\right)\\
p_5 &= p(a)\,p(b')\,q(AB')\\
p_6 &= p(a)\,p(b)\,q(AB)\\
p_7 &= p(a')\,p(b)\left(q(B)-q(A'B)\right)\\
p_8 &= p(a')\,p(b)\left(1-q(A')-q(B)+q(A'B)\right)\\
p_9 &= p(a')\,p(b)\left(q(A')-q(A'B)\right)\\
p_{10} &= p(a')\,p(b')\left(q(A')-q(A'B')\right)\\
p_{11} &= p(a')\,p(b')\,q(A'B')\\
p_{12} &= p(a')\,p(b')\left(1-q(A')-q(B')+q(A'B')\right)\\
p_{13} &= p(a')\,p(b')\left(q(B')-q(A'B')\right)\\
p_{14} &= p(a)\,p(b')\left(q(B')-q(AB')\right)\\
p_{15} &= p(a)\,p(b)\left(q(B)-q(AB)\right)\\
p_{16} &= p(a')\,p(b)\,q(A'B)
\end{aligned}
\tag{22}
$$

Probabilities $p_1, p_2, \ldots p_{16}$ unambiguously determine the probability function on the entire algebra. Denote $(\mathcal{A}, p)$ the probability model we obtained.

**10.** The extension will be realized gradually, in four steps[14]. Let $(X,Y)$ be one of the correlated pairs $(A,B)$, $(A,B')$, $(A',B)$, $(A',B')$. As a first step, algebra $(\mathcal{A}, p)$ will be extended such that the extension contains a common cause for the correlation of $(X,Y)$. I would like to emphasize that common cause means now an event satisfying the more restrictive conditions (10)-(19), instead of the original definition (1)-(5).

As an ansatz, assume that $p\left(X \mid Z_{XY}\right)=\xi_X\, p(x)$ and $p\left(Y \mid Z_{XY}\right)=\xi_Y\, p(y)$, where $\xi_X$ and $\xi_Y$ are parameters independent of $p(x)$ and $p(y)$. According to the theorem of total probability, one can easily show that probabilities

[14]Some elements of the calculation in following points **10.-12.** are similar to the Proof of Theorem 3. in Hofer-Szabó, G., Rédei, M., Szabó, L. E., On Reichenbach's common cause principle and Reichenbach's notion of common cause, *http://xxx.lanl.gov/ quant-ph/9805066*. I must, however, draw the reader's attention to the principal discrepancy between the two cases. The difference comes from the divergence between the concepts of common cause in (1)-(5) and in (10)-(19) , namely, that only the later can be applied to the EPR-Aspect scenario. In the above-mentioned paper we proved that every classical Kolmogorovian probability space as well as every quantum probability model can be enlarged by finite number of arbitrary non-common common causes satisfying conditions (1)-(5). Unfortunately, none of these results are applicable in the EPR-Aspect context: the reason is that in the classical case common cause is understood as in (1)-(5), therefore quantities $q(A), q(B), \ldots$ should be interpreted as classical probabilities, which is impossible. In the quantum case $q(A), q(B), \ldots$ are "quantum probabilities" and the common cause is a "quantum event", that means the whole construction remains only a formal, mathematical (although highly non-trivial) one, because of the very same reason I put forward in point **4.**

$$
\begin{aligned}
p\left(Z_{XY}\right) &= \frac{\Delta_q\left(XY\right)}{\left(q\left(X\right)-\xi_X\right)\left(q\left(Y\right)-\xi_Y\right)+\Delta_q\left(XY\right)} \\
p\left(X \mid Z_{XY}\right) &= \xi_X p\left(x\right) \\
p\left(Y \mid Z_{XY}\right) &= \xi_Y p\left(y\right) \\
p\left(X \mid \overline{Z}_{XY}\right) &= \frac{q\left(XY\right)-q\left(X\right)\xi_Y}{q\left(Y\right)-\xi_Y} p\left(x\right) \\
p\left(Y \mid \overline{Z}_{XY}\right) &= \frac{q\left(XY\right)-q\left(Y\right)\xi_X}{q\left(X\right)-\xi_X} p\left(y\right)
\end{aligned}
$$

satisfy equations (17)-(19), whenever $\xi_X$ and $\xi_Y$ satisfy the following conditions:

$$
\begin{array}{ll}
\left\{\begin{bmatrix} 1 \geq \xi_X \geq \dfrac{q(XY)}{q(Y)} \\ 1 \geq \xi_Y \geq \dfrac{q(XY)}{q(X)} \end{bmatrix} \text{ or } \begin{bmatrix} 0 \leq \xi_X \leq \dfrac{q(X)-q(XY)}{1-q(Y)} \\ 0 \leq \xi_Y \leq \dfrac{q(Y)-q(XY)}{1-q(X)} \end{bmatrix}\right\} & \text{if } \Delta_q(XY) > 0 \\
\left\{\begin{bmatrix} 0 \leq \xi_X \leq \dfrac{q(XY)}{q(Y)} \\ 1 \geq \xi_Y \geq \dfrac{q(Y)-q(XY)}{1-q(X)} \end{bmatrix} \text{ or } \begin{bmatrix} 1 \geq \xi_X \geq \dfrac{q(X)-q(XY)}{1-q(Y)} \\ 0 \leq \xi_Y \leq \dfrac{q(XY)}{q(X)} \end{bmatrix}\right\} & \text{if } \Delta_q(XY) < 0
\end{array}
\tag{23}
$$

Thus, parameters $\xi_X$ and $\xi_Y$ together with inequalities (23) make a complete survey of the possible values of probabilities $p\left(Z_{XY}\right)$, $p\left(X \mid Z_{XY}\right)$, $p\left(Y \mid Z_{XY}\right)$, $p\left(X \mid \overline{Z}_{XY}\right)$, $p\left(Y \mid \overline{Z}_{XY}\right)$.

Now we are going to show that for any admissible values of parameters $\xi_X$ and $\xi_Y$ there is an extension of the event algebra, such that it contains a common cause corresponding to these parameters.

**11.** In order to obtain an extension, let us “duplicate” the atoms of the original algebra. That is, consider two copies of set $\Omega$ :

$$
\Omega_i \overset{def}{=} \left\{(u,i) \mid u \in \Omega\right\} \quad (i=1,2)
$$

Let $h_i$ $(i=1,2)$ be the two corresponding algebra homomorphisms:

$$
h_i : U \in \mathcal{A} \mapsto \left\{(u,i) \mid u \in U\right\} \in \mathcal{A}_i = 2^{\Omega_i} \quad (i=1,2)
$$

Let the extended algebra be $\tilde{\mathcal{A}} = 2^{\Omega_1 \cup \Omega_2}$. Obviously, every element of $\tilde{\mathcal{A}}$ can be imagined in form of $h_1(U) \cup h_2(V)$, where $U,V \in \mathcal{A}$. The original probability model can be embedded by mapping $h : U \in \mathcal{A} \mapsto h(U) = h_1(U) \cup h_2(U) \in \tilde{\mathcal{A}}$, with the requirement that

$$
p(U) = \tilde{p}\left(h(U)\right) \quad \left(\forall\, U \in \mathcal{A}\right) \tag{24}
$$

where $\tilde{p}$ is the probability function defined on $\tilde{\mathcal{A}}$.

Condition (24) can be satisfied if probability function $\tilde{p}$ is defined in the following way:

$$
\begin{aligned}
\tilde{p}\left(h_1(U)\cup h_2(V)\right) &\stackrel{def}{=} \lambda_1 p\left(Ux\overline{(X\cup y)}\right)+\lambda_2 p\left(UX\overline{y}\right)+\lambda_3 p\left(Uxy\overline{(X\cup Y)}\right)\\
&+\lambda_4 p\left(UyX\overline{Y}\right)+\lambda_5 p\left(UXY\right)+\lambda_6 p\left(UYx\overline{X}\right)+\lambda_7 p\left(Uy\overline{(x\cup Y)}\right)\\
&+\lambda_8 p\left(UY\overline{x}\right)+\lambda_9 p\left(U\overline{(x\cup y)}\right)+\left(1-\lambda_1\right)p\left(Vx\overline{(X\cup y)}\right)\\
&+\left(1-\lambda_2\right)p\left(VX\overline{y}\right)+\left(1-\lambda_3\right)p\left(Vxy\overline{(X\cup Y)}\right)+\left(1-\lambda_4\right)p\left(VyX\overline{Y}\right)\\
&+\left(1-\lambda_5\right)p\left(VXY\right)+\left(1-\lambda_6\right)p\left(VYx\overline{X}\right)+\left(1-\lambda_7\right)p\left(Vy\overline{(x\cup Y)}\right)\\
&+\left(1-\lambda_8\right)p\left(VY\overline{x}\right)+\left(1-\lambda_9\right)p\left(V\overline{(x\cup y)}\right)
\end{aligned}
\tag{25}
$$

where

$$
\begin{aligned}
\lambda_1 &= \frac{\left(1-\xi_X\right)}{1-q(X)}\cdot\frac{\Delta_q(XY)}{\left(q(X)-\xi_X\right)\left(q(Y)-\xi_Y\right)+\Delta_q(XY)}\\
\lambda_2 &= \frac{\xi_X}{q(X)}\cdot\frac{\Delta_q(XY)}{\left(q(X)-\xi_X\right)\left(q(Y)-\xi_Y\right)+\Delta_q(XY)}
\end{aligned}
\tag{26}
$$

$$
\begin{aligned}
\lambda_3 &= \frac{\left(1-\xi_X-\xi_Y+\xi_X\xi_Y\right)}{1-q(X)-q(Y)+q(XY)}\cdot\frac{\Delta_q(XY)}{\left(q(X)-\xi_X\right)\left(q(Y)-\xi_Y\right)+\Delta_q(XY)}\\
\lambda_4 &= \frac{\left(\xi_X-\xi_X\xi_Y\right)}{q(X)-q(XY)}\cdot\frac{\Delta_q(XY)}{\left(q(X)-\xi_X\right)\left(q(Y)-\xi_Y\right)+\Delta_q(XY)}\\
\lambda_5 &= \frac{\xi_X\xi_Y}{q(XY)}\cdot\frac{\Delta_q(XY)}{\left(q(X)-\xi_X\right)\left(q(Y)-\xi_Y\right)+\Delta_q(XY)}\\
\lambda_6 &= \frac{\left(\xi_Y-\xi_X\xi_Y\right)}{q(Y)-q(XY)}\cdot\frac{\Delta_q(XY)}{\left(q(X)-\xi_X\right)\left(q(Y)-\xi_Y\right)+\Delta_q(XY)}\\
\lambda_7 &= \frac{\left(1-\xi_Y\right)}{1-q(Y)}\cdot\frac{\Delta_q(XY)}{\left(q(X)-\xi_X\right)\left(q(Y)-\xi_Y\right)+\Delta_q(XY)}\\
\lambda_8 &= \frac{\xi_Y}{q(Y)}\cdot\frac{\Delta_q(XY)}{\left(q(X)-\xi_X\right)\left(q(Y)-\xi_Y\right)+\Delta_q(XY)}\\
\lambda_9 &= \frac{\Delta_q(XY)}{\left(q(X)-\xi_X\right)\left(q(Y)-\xi_Y\right)+\Delta_q(XY)}
\end{aligned}
\tag{26}
$$

This is so because the elements $x\overline{(X\cup y)}$, $X\overline{y}$, $xy\overline{(X\cup Y)}$, $yX\overline{Y}$, $XY$, $Yx\overline{X}$, $y\overline{(x\cup Y)}$, $Y\overline{x}$ and $\overline{(x\cup y)}$ form a complete disjoint system in $\mathcal{A}$. (See **Figure 3**.)

It remains to be told which element of the extended algebra is the common cause: $Z_{XY}=\Omega_1\in\tilde{\mathcal{A}}$. It can be easily verified, using (23), (25) and (26), that

$$
\tilde{p}\left(Z_{XY}\right)=\tilde{p}\left(\Omega_1\right)=\frac{\Delta_q(XY)}{\left(q(X)-\xi_X\right)\left(q(Y)-\xi_Y\right)+\Delta_q(XY)}
\tag{27}
$$

as it is expected, taking into account what was said in point **10.**

Of course, the above construction can be meaningful only if it is true that $\lambda_1, \lambda_2, \ldots \lambda_9 \in [0, 1]$. It is easily verifiable that, fortunately, this is true for all possible $\xi_X, \xi_Y$ satisfying conditions (23), if the conditional probabilities are equal to the values given in (8).

**12.** The probability model $(\tilde{\mathcal{A}}, p)$ we obtained, via embedding $h$, contains 1) all the original events with the original probabilities and, consequently, with the original correlations, and 2) a new event that is a common cause for the correlation of the selected pair $(X, Y)$, satisfying conditions (10)-(19).

Repeating the above procedure another three times we obtain an extension with 256 atoms, that contains common causes for all of the four correlations. The common causes thus gained $Z_{AB}$, $Z_{AB'}$, $Z_{A'B}$, $Z_{A'B'}$ are different, of course, in accordance with the violation of the Clauser-Horne inequality.

## *The charge of "world conspiracy"*

**13.** Notice that the probabilities of the atoms in (22) are not independent of the probabilities of the measurement choices. This fact might awaken suspicion that I suggest a solution for the EPR-Bell problem, which implies a strange "conspiracy" in our world. In some of my earlier papers,[15] I shown deterministic hidden parameter models for the EPR-Aspect experiment, that, indeed, embraced an element of conspiracy: the choice of the measurement was not independent of the common cause. Now the situation is entirely different. *The common cause satisfies condition* (13), *which is just destined for expressing the independence between the measurement choices and the common cause.*

This raises, however, the following two questions: 1) Why is it so that the probabilities of the atoms of the event algebra depend on the probabilities of the measurement choices? 2) How can this common cause model be accommodated in a relativistic world? I will give a joint answer to these questions by considering how can we describe the fundamentally stochastic EPR-Aspect experimental scenario within the relativistic and deterministic framework.

**14.** Assume that our world is deterministic and relativity theory holds. Any stochastic feature of such a world is of epistemic origin, that is, it can be derived from the fact that some of the determinant data are unknown to us. At the same time, it is true that everything what is going on in a spacetime region is determined by the causal past of that region. Intertwining these two ideas, with high generality we may assume that the spacetime patterns of the repeated runs of the experiment are such as it is shown in **Figure 3** $D_1, D_2, \ldots$ are the domains of dependence of hyper-surfaces $S_1, S_2, \ldots$. The "state of affairs" along a hype-surface unambiguously determines what is going on in the corresponding domain. These determinant data, however, are not completely known: There are "hard" data, which are known, and which are the same on each hyper-surface, and on the basis of which we regard the "stories" going on in the different domains $D_1, D_2, \ldots$ as repetitions of the same experimental scenario. While, there are "soft" parameters along the hyper-surfaces, which are unknown to (or neglected by) us, and which are presumably of different value along the hyper-surfaces $S_1, S_2, \ldots$. These hyper-surfaces are, nevertheless, regarded as the representatives of the same physical condition. The statistical ensemble consists of these $(S_i, D_i)$ patterns. We can measure the probability of an event observed in the experiment by counting the relative frequency that how many times does such an $(S, D)$-pattern appear, which contains the corresponding event.

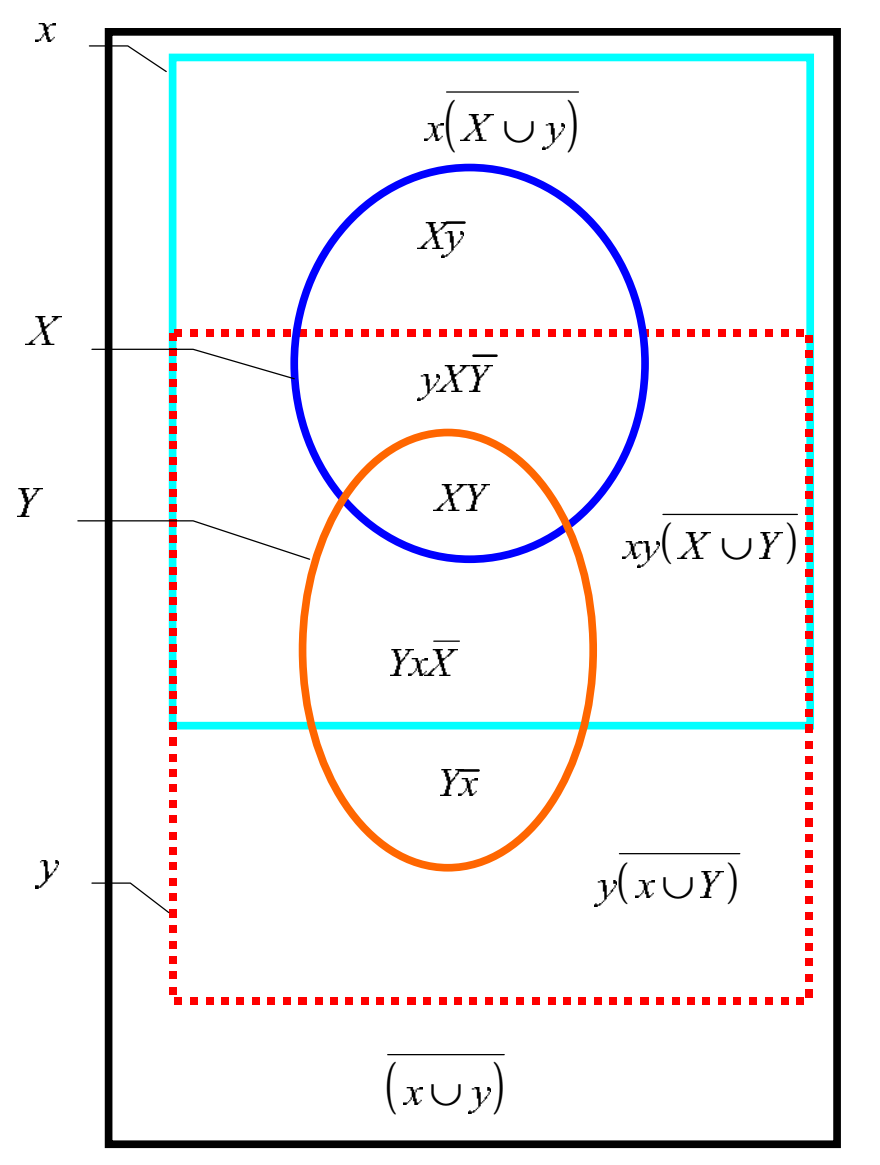

***Figure 3*** $x\overline{(X\cup y)}$, $X\overline{y}$, $xy\overline{(X\cup Y)}$, $yX\overline{Y}$, $XY$, $Yx\overline{X}$, $y\overline{(x\cup Y)}$, $Y\overline{x}$, $\overline{(x\cup y)}$ form a complete disjoint system in $\mathcal{A}$.

[15] Szabó, L. E., Is quantum mechanics compatible with a deterministic universe? Two interpretations of quantum probabilities, *Foundations of Physics Letters*, **8**, 421-440. (1995); Quantum Mechanics in an Entirely Deterministic Universe, *Int. J. Theor. Phys.* **34**, 1751-1766. (1995)

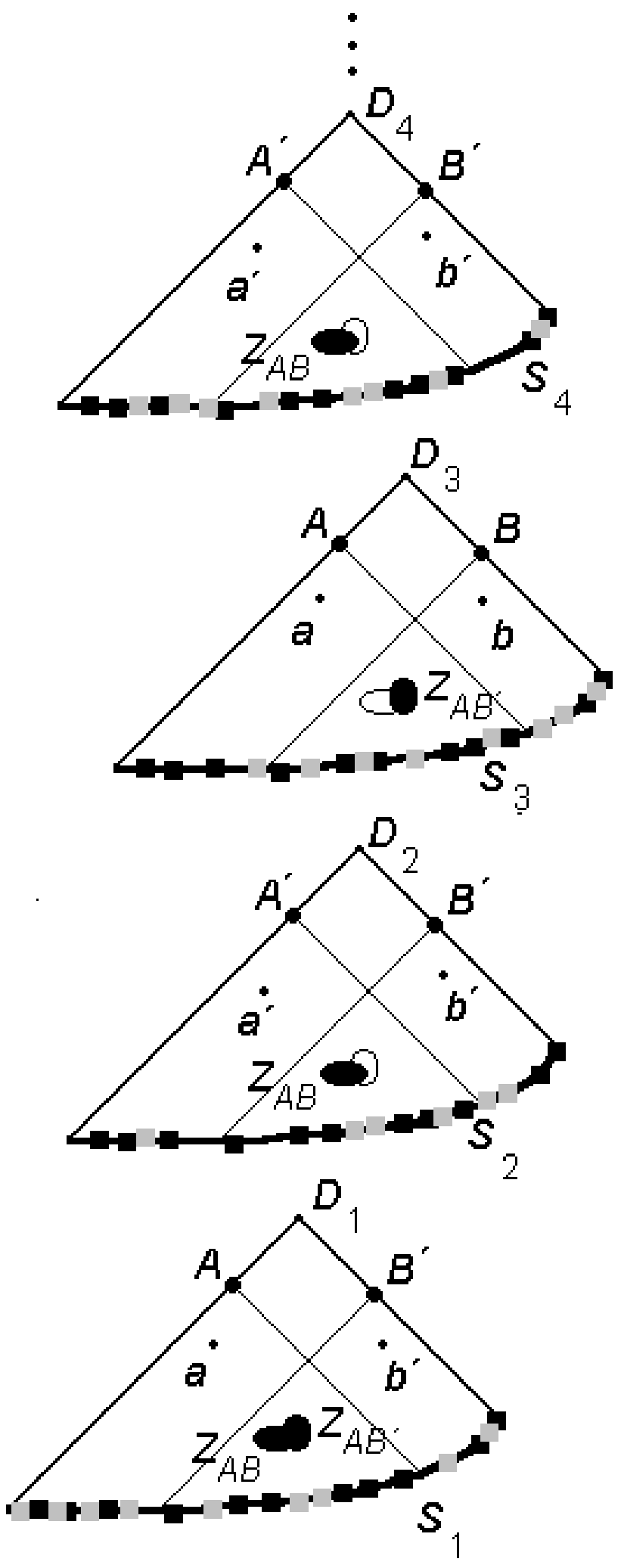

*Figure 4* $D_1, D_2, \ldots$ *are the domains of dependence of hyper-surfaces* $S_1, S_2, \ldots$. *The "state of affairs" along a hype-surface unambiguously determines what is going on in the corresponding domain. These determinant data, however, are not completely known: ■ denotes the "hard" data, which are known, while ■ stands for the unknown "soft" data.*

Nothing excludes that – as it is just expected on the basis of relativistic considerations – the common cause lies on the intersection of the past light-cones of the two measurement outcomes, since it is correlated with both of them. At the same time, the measurement choices may be completely independent of the common cause, since there is no correlation among them. It is just possible, for instance, that the choices of the measurements are governed by signals coming from outside of the intersection of the past light-cones (e.g. from a far distant part of the universe).

An atom of the event algebra is nothing else but a possible type, or sub-class of the $(S, D)$-patterns, and as such, its relative frequency obviously cannot be independent of the relative frequencies of the measurement choices.

Finally, I have to emphasize one more "no-conspiracy" feature of the common cause model we created: as we can see from **Figure 4**, the common cause is not only statistically independent of the measurement choices, but *its probability too is independent of the probabilities of the measurement choices.*

Thus, the common cause model of the EPR-Aspect experiment, the existence of which we have proven, can be completely compatible with relativity theory (and determinism) and does not imply “world conspiracy” at all.

## *Conclusions*

**15.** The EPR-Bell paradox consists in the problem that we observe correlations between spatially separated events in the EPR-experiments, which do not admit common-cause-type explanation; and it must therefore be inevitably concluded, that, contrary to relativity theory, in the realm of quantum physics there exists action at a distance, or at least superluminal causal propagation is possible; that is, either relativity theory or Reichenbach’s common cause principle fails. In this paper it was explicitly shown that the correlations we encounter in the EPR experiment could have common causes. Moreover, these common causes are compatible with locality principle. This claim is bold enough to point out again what is that novelty in the present approach, that enables us to resolve the paradox. This will be briefly sketched in the following points:

The most important recognition is that “quantum probability” must not be interpreted as probability. It has been clear for a long time that such an interpretation is problematic, if we insist to the frequentists’ understanding of probability. The quantum mechanical $\mathrm{tr}\left(\hat{W}\hat{A}\right)$ is not the (absolute) probability of a real event, but it is a *conditional* probability $p(A \mid a)$, which means the probability of the outcome-event $A$, given that the measurement-preparation $a$ has happened. $\mathrm{tr}\left(\hat{W}\hat{A}\right)$ no doubt does have such a meaning, I believe, in accordance with every day’s laboratory practice. That is why one rightly calls this interpretation as “minimal” interpretation. The controversial question is, whether it means something more: whether there exists an event $\tilde{A}$ in reality, such that $\mathrm{tr}\left(\hat{W}\hat{A}\right) = p\left(\tilde{A}\right)$ holds. It is the Pitowsky theorem which answers this question: If probability means relative frequency, then the probability function on an event algebra can be nothing else but the weighted average of the classical truth-value functions, therefore, according to the theorem, it must be Kolmogorovian, consequently, it satisfies the Bell inequalities. Since the $\mathrm{tr}\left(\hat{W}\hat{A}\right)$-type quantities violate Bell inequalities, they are not interpretable as relative frequencies. In other words, there are no events that would happen with probabilities like $\mathrm{tr}\left(\hat{W}\hat{A}\right)$. And accordingly, as it was pointed out in point **4.**, the question about the existence of common cause, in its original form, is meaningless, because it is about the existence of common cause for correlations among non-existing events. *What the violation of the Bell inequalities indicates is not that correlations (9) do not have common cause, but rather that there are no events which would have such correlations.*

That is why we had to reformulate the problem of existence of common cause, only with reference to events occurring in reality and to their factual relative frequencies. These relative frequencies do satisfy Pitowsky’s Bell inequalities. In this new formulation, too, we could derive a Bell inequality (more precisely, a Clauser-Horne inequality), the meaning of which was not, however, the same as that of the inequality derived by Pitowsky, because it applied to conditional probabilities instead of absolute ones. The conditional probabilities calculated from the observed relative frequencies in the EPR-Aspect experiment violate this inequality, from which we might already correctly infer the failure of the CCP in quantum mechanics.

In the new derivation of the Clauser-Horne inequality, however, we explicitly stipulated an assumption that has been tacitly included in the concept of common cause before, namely that all correlations in the experiment have one single common common cause. The CCP does not require this, and there is no particular reason in case of the EPR-Aspect experiment either to insist on this unjustified condition.

The last but very important step in resolving the paradox was the realization that the fact in itself that the measurement choices were taken into consideration does not necessarily led to a "world conspiracy"[16]. It is quite natural, that the atoms of the event algebra, which can be identified with the various "states of affairs" or truth-value functions, are not independent of the measurement choices. From this circumstance, however, it dos not follow that there couldn't exist common causes satisfying Bell's requirement[17] of independence between the common causes and the measurement choices. Moreover, we actually proved the existence of such common cause events, by constructing them explicitly. In addition, each common cause and the measurement choices are independent not only statistically, but also in the sense that independently of the frequencies with which the different measurements are chosen, 1) the common cause is always a common cause, 2) it is always the same element of the event algebra, and finally, 3) the probability of the occurrence of the common cause is always the same. All these are in accordance with the presumable temporal order of the events in question: For sake of simplicity assume that the common cause temporally precedes the choices of the measurements. The event algebra necessarily contains events occurring at different times. Therefore, of course, the probability of the atoms (the relative frequencies of the possible truth-value functions) depend on the measurement choices, too, which happen later than the common cause. At the same time, the probability of the common cause event, occurring earlier, cannot depend on the later choices of measurements.

Our common cause model of the EPR-Aspect experiment does satisfy all conditions having been required in the EPR-Bell literature.

## *Postscript*

A few weeks after the Internet publication of this preprint I realized a new problem with the above attempt to solve the EPR problem via the Reichenbachian concept of common cause. Instead of removing the paper from the archives I would like to share this problem with the occasional readers.

The problem is this: While it is true that each common cause event is independent of the measurement choices, and therefore, the model, indeed, satisfies all locality conditions having been required in the EPR-Bell literature, it turns out that such combinations of the common cause events as

$$\begin{gathered} Z_{AB} \wedge Z_{AB'} \\ Z_{AB} \vee Z_{AB'} \\ Z_{AB} \wedge Z_{AB'} \wedge Z_{A'B} \\ etc. \end{gathered}$$

do statistically correlate with the measurement operations.

First I hoped that I can modify the model in order to remove these unwanted correlations, but I couldn't. On the contrary, a whole series of computer models convinced me that the task is hopeless: the numerical calculations show that there is no such a modification of the model, that the above mentioned events too are statistically independent of the measurement choices. However, the exact answer is not known yet.

[16] We don't need—at least not with the aim of resolving the EPR-Bell paradox—to accept such a bizarre metaphysical assumption as Huw Price's "backward causation" either. Cf., Price, H., A Neglected Route to Realism about Quantum Mechanics, in Grim, P., Mar, G. and Williams, P. (eds.) *The Philosopher's Annual*, Vol. XVII, Ridgeview, 1996, 181-215. (http://xxx.lanl.gov/abs/gr-qc/9406028)

[17] Bell, J. S., *Speakable and unspeakable in quantum mechanics. Collected papers on quantum philosophy*, Cambridge University Press, Cambridge, 1987, p. 154.